\documentclass[reprint,longbibliography,
nofootinbib,
amsmath,amssymb,
aps,
pra,
floatfix,
showpacs
]{revtex4-2}

\usepackage{graphicx}
\usepackage{dcolumn}
\usepackage{bm}

\usepackage{natbib}
\usepackage[colorlinks]{hyperref}
\hypersetup{linkcolor=red,citecolor=blue,urlcolor=blue}

\begin{document}

\title{Vector interactions inhibit quark-hadron mixed phases in neutron stars} 

\author{G. Lugones$^{1}$ and A. G. Grunfeld$^{2,3}$}

\affiliation{$^1$ Universidade Federal do ABC, Centro de Ci\^encias Naturais e Humanas, Avenida dos Estados 5001- Bang\'u, CEP 09210-580, Santo Andr\'e, SP, Brazil.\\
$^2$CONICET, Godoy Cruz 2290, Buenos Aires, Argentina \\
$^3$Departamento de F\'\i sica, Comisi\'on Nacional de Energ\'{\i}a At\'omica, Av. Libertador 8250, (1429) Buenos Aires, Argentina}

\begin{abstract}
We investigate the surface tension $\sigma$ and the curvature energy $\gamma$ of quark matter drops in the MIT bag model with vector interactions. Finite size corrections to the density of states are implemented by using the multiple reflection expansion (MRE) formalism. We find that $\sigma$ and $\gamma$ are strongly enhanced by new terms arising from vector interactions. With respect to the noninteracting case they are increased by a large factor,  which can be as high as $\sim 10$ when the vector coupling constant $g$ varies within the range used in the literature. This behavior may have major  consequences for the hadron-quark mixed phase speculated to exist at neutron star (NS) interiors, which may be totally suppressed or have its extension substantially reduced.
\end{abstract}

\maketitle

\section{Introduction}

It has been speculated that hybrid stars may contain a  hadron-quark mixed phase in their interiors. In such a phase, the electric charge is zero globally but not locally, i.e. both the hadronic and the quark phases have net charges, but the whole mixture is electrically neutral. 
The mixed phase is usually studied  in the Wigner-Seitz approximation,  where the whole space is divided into equivalent periodically  repeating charge-neutral cells with given geometrical symmetry, which may change from droplet to rod, slab, tube, and bubble with increasing baryon density \cite{Glendenning:2001pe}. Within each cell, a lump made of one phase is embedded in the other one and both share a common lepton background. Both phases are separated by a sharp boundary at which it is required chemical, mechanical and thermal equilibrium between them. 

The existence of the mixed phase depends crucially on the amount of electrostatic and surface energy needed for the formation of geometric structures along a wide range of densities \cite{Voskresensky:2002hu,Endo:2011em,Yasutake:2014oxa}. If the energy cost of Coulomb and surface effects exceeds the gain in bulk energy, the mixed phase turns out to be energetically disfavored with respect to a simple sharp interface between locally neutral hadronic and locally neutral quark matter.

Our goal in this work is to determine the surface tension and curvature energy of quark drops in the mixed phase in order to assess whether its existence is energetically favored inside NSs. 
Several works in the literature have used the surface tension as a free parameter to describe the possible structure of mixed phases, and to evaluate under which conditions it would be favored over a sharp quark-hadron interface (see e.g. \cite{Voskresensky:2002hu,Alford:2006bx,Maslov:2018ghi,Yasutake:2014oxa,Wu:2017xaz,Wu:2018zoe,Xia:2020brt} and references therein). However, there are fewer works that have determined $\sigma$ and $\gamma$ from microscopic calculations. 
Although there are some works evaluating $\sigma$ for vanishing chemical potentials based on lattice QCD  \cite{Huang:1990jf, Alves:1992xv, Brower:1992bm, deForcrand:2005pb, deForcrand:2004jt}, such approach is not possible  for dense matter, where one depends on effective models. 
Calculations made within the MIT bag model \cite{Berger:1986ps,Lugones:2016ytl,Lugones:2018qgu,Lugones:2020qll,Ju:2021hoy},  the Nambu-Jona-Lasinio model \cite{Garcia:2013eaa, Ke:2013wga},  the linear sigma model \cite{Palhares:2010be, Pinto:2012aq, Kroff:2014qxa},  the three-flavor Polyakov-quark-meson model \cite{Mintz:2012mz}, the Dyson-Schwinger equation approach~\cite{Gao:2016hks},  the nucleon-meson model~\cite{Fraga:2018cvr}, and the equivparticle model~\cite{Xia:2018wdj}, predict small values of the surface tension, typically below $30 \mathrm{MeV/fm}^{2}$. Somewhat higher values, in the range $\sigma= 30-70 \mathrm{MeV/fm}^{2}$, are obtained within the quasiparticle model \cite{Wen:2010zz}. Significantly larger results ($\sigma= 145 - 165 \mathrm{MeV/fm}^{2}$) arise within the Nambu-Jona-Lasinio model when the MRE method is used model~\cite{Lugones:2013ema}. 

In the context of phenomenological models, interest in repulsive vector interactions has reemerged in recent years (see e.g. \cite{Lenzi:2012xz, Klahn:2015mfa, Dexheimer:2018dhb, Alvarez-Castillo:2018pve, Salinas:2019fmu,  Ayriyan:2021prr, Lopes2021_I, Lopes2021_II,Benic:2014jia,Kaltenborn:2017hus} and references therein) because they help explaining large observed NS masses \cite{Demorest:2010bx, Antoniadis:2013pzd, Cromartie:2019kug, Riley:2021pdl, Miller:2021qha}. To the best of our knowledge, a detailed and self-consistent analysis of the role of vector interactions in $\sigma$ and $\gamma$ of astrophysical quark matter has not been presented in the literature.   In this work we analyze this problem and identify new vector contributions to the expressions for $\sigma$ and $\gamma$. We will show that these quantities are strongly enhanced by vector interactions, which has significant  consequences for the internal structure of NSs.

\section{The vector MIT bag model in bulk}

Quark matter is described by the MIT bag model with vector interactions, which are introduced by a vector-isoscalar meson $V^{\mu},$ with coupling constants $g_{qqV},$ coupling to all three quarks. The Lagrangian density of the model reads \cite{Franzon:2016urz,Lopes2021_I,Lopes2021_II}:
\begin{equation}
\begin{aligned}
\mathcal{L} & = \sum_{q}\left\{\bar{\psi}_{q}\left[i \gamma^{\mu} \partial_{\mu}-m_{q}\right] \psi_{q}-B\right\} \Theta\left(\bar{\psi}_{q} \psi_{q}\right)  \\
& + \sum_{q} g_{q q V}\left\{\bar{\psi}_{q}\left[\gamma^{\mu} V_{\mu}\right] \psi_{q}\right\} \Theta\left(\bar{\psi}_{q} \psi_{q}\right) +   \tfrac{1}{2} m_{V}^{2} V_{\mu} V^{\mu} \\
& + \sum_{l} \bar{\psi}_{l} \gamma_{\mu}\left(i \partial^{\mu}-m_{l}\right) \psi_{l}   ,
\end{aligned}
\end{equation}
where $q$ runs over quarks ($u$, $d$ and $s$), $l$ over leptons ($e^{-}$ and $\nu_e$),  the bag constant $B$ represents the extra energy per unit of volume required to create a region of perturbative vacuum \cite{Farhi:1984qu}, and  $\Theta$ is the Heaviside step function  ($\Theta=1$ inside the bag; $\Theta=0$ outside).  
For simplicity we adopt here a universal coupling of the quark $q$ with the vector field $V^{\mu}$, i.e. the coupling constants verify $g_{s s V}=g_{u u V}=g_{d d V} \equiv g$. The mass of the vector field is taken to be $m_{V} = 780 \; \mathrm{MeV}$.

Working in the mean field approximation (MFA) and defining $G_{V} \equiv \left({g}/{m_{V}}\right)^{2}$,  the  equation for the vector field reads:
\begin{equation}
m_{V} V_{0}  = G_V^{1/2} (n_u + n_d + n_s),  
\end{equation}
where $n_{q}=\left\langle\bar{\psi}_{q} \gamma^{0} \psi_{q}\right\rangle$ is the  quark number density\footnote{In the MFA vector MIT model considered here, negative values of $G_V$ are not allowed.  In fact, the vector terms are the same as in the Walecka model \cite{Walecka:1974qa} where the vector mean field turns out to be repulsive. This is apparent from the dispersion relation $E_q= (m_q^2 + k^2)^{1/2} + g V^0$ \cite{Lopes2021_II}, since the vector contribution $g V^{0}$ is positive in the MFA. This is so because  $g V^{0}$ is given by $g V_{0} =(g/m_V)^2 \sum_q n_q$, and the right hand side of the equation is positive definite. This means that not only $G_V \equiv (g/m_V)^2 > 0$ but also $G_V^{1/2} \equiv g/m_V >0$.}.
The grand thermodynamic potential per unit volume is \cite{Lopes2021_II}:
\begin{eqnarray}
\Omega & =&\sum_{i = q, l} \Omega^*_{i} + B-\tfrac{1}{2} m_{V}^{2} V_{0}^{2}
\label{eq:Omega_bulk}
\end{eqnarray}
where
\begin{equation}
\Omega^*_{i}= - \frac{g_{i}}{6 \pi^{2}} \int_{0}^{\infty} \frac{(f^*_{i+} + f^*_{i-} ) }{\sqrt{k^2 + m_{i} ^2}}  k^4 d k, 
\end{equation}
being $k$ the particle's momentum and $g_i$ a degeneracy factor ($g_q=6$, $g_{e^-}=2$, $g_{\nu_e}=1$). The Fermi--Dirac distribution functions for particles and antiparticles are:
\begin{equation}
f^*_{i \pm}=\frac{1}{1+\exp \left[\left((k^2 + m_i^2)^{1/2} \mp \mu_{i}^{*}\right) / T\right]} .
\label{eq:effective_FD}
\end{equation}
For quarks, the effective chemical potential reads:
\begin{eqnarray}
\mu_{q}^{*} &=&  \mu_{q} -  G_V^{1/2} m_V V_{0} + q_q |e \phi| 
\label{eq:effective_mu}
\end{eqnarray}
where we added the contribution of an  electrostatic potential $\phi$,  being $e$ the electron's electric charge, $q_u=2/3$, and $q_d = q_s= -1/3$. For $e^{-}$ we have $\mu_{e}^{*}=  \mu_{e} - |e \phi|$. $\mu_{i}$ is the chemical potential of the $i$-species.
The particle number density of each species is: 
\begin{eqnarray}
n_{i} =  \frac{g_{i}}{2 \pi^2} \int_{0}^{\infty} (f^*_{i+}- f^*_{i-}) k^2 dk .
\end{eqnarray}
%

\section{Finite size effects with vector interactions}

Effects due to the finite size of quark droplets will be taken into account within the MRE framework  \cite{Balian:1970fw,Madsen:1994vp,Kiriyama:2002xy,Kiriyama:2005eh}. 
The basic idea of the MRE, is that the propagation of a particle in a cavity can occur either directly (as described by the free space propagator $S^0$), or via one or more reflections at the surface, as can be seen in Fig. 1 of Ref. \cite{Hansson:1983xt}. The expansion for the time independent Green’s function reads
\begin{equation}
\begin{aligned} S\left(\mathbf{r}, \mathbf{r}^{\prime}\right) =S^{0}\left(\mathbf{r}, \mathbf{r}^{\prime}\right)+\oint_{\partial \Omega} d \sigma_{\alpha} S^{0}(\mathbf{r}, \boldsymbol{\alpha}) K(\boldsymbol{\alpha}) S^{0}\left(\boldsymbol{\alpha}, \mathbf{r}^{\prime}\right) \\ +\oint_{\partial \Omega} d \sigma_{\alpha} d \sigma_{\beta} S^{0}(\mathbf{r}, \boldsymbol{\alpha}) K(\boldsymbol{\alpha}) S^{0}(\boldsymbol{\alpha}, \boldsymbol{\beta}) K(\boldsymbol{\beta}) S^{0}(\boldsymbol{\beta}, \mathbf{r}^{\prime})  + ... \end{aligned}
\end{equation}
being $K$ a reflection kernel describing the interplay with the surface due to the confining boundary conditions \cite{Hansson:1982cu}.

The density of states $\rho$ is obtained replacing the latter general form for $S\left(\mathbf{r}, \mathbf{r}^{\prime}\right)$ in the trace formula  $\rho(\omega) =\mp \frac{1}{\pi} \operatorname{Im} \operatorname{tr} S(\omega \pm i \varepsilon) \gamma^{0}$   \cite{Jensen:1996uku}:
\begin{equation}
\begin{aligned} 
\rho(\omega) = 
& \mp \frac{1}{\pi} \operatorname{Im} \operatorname{tr} S^{0}(\omega \pm i \varepsilon) \gamma^{0} \\
& \mp \frac{1}{\pi} \int_{\Omega} d^{3} \mathbf{r} \oint_{\partial \Omega} d \sigma_{\alpha} \lim _{\mathbf{r}^{\prime} \rightarrow \mathbf{r}} \operatorname{Im} S^{0}(\mathbf{r}, \boldsymbol{\alpha}, \omega \pm i \varepsilon) \\ &\times K(\boldsymbol{\alpha}) S^{0}(\boldsymbol{\alpha}, \mathbf{r}^{\prime}, \omega \pm i \varepsilon) \gamma^{0}  +... \end{aligned}
\end{equation}   
The first term gives the volume contribution to $\rho$, and from the second one, terms proportional to the surface area $S$ and the extrinsic curvature $C$ can be extracted. As a consequence, for spherical drops one obtains: 
\begin{equation}
\rho_{i}(k) = 1+\frac{6 \pi^{2}}{k R} f_{S,i}+\frac{12 \pi^{2}}{(k R)^{2}} f_{C,i} ,
\label{rho_MRE}
 \end{equation}
which is the same for quarks as for antiquarks.  The volume term is independent of the boundary condition and of the type of field (scalar, spinor, vector, etc.), but the functions $f_S$ and $f_C$ depend on the boundary condition as well as on the nature of the field. A detailed calculation of $f_S$ and $f_C$ for quarks has never been published, but the result for $f_S$  was given in \cite{Berger:1986ps,Mardor:1991dt}:
\begin{equation}
f_{S,i}(k) = - \frac{1}{8 \pi} \left(1 - \frac{2}{\pi} \arctan \frac{k}{m_i} \right)  .
\label{eq:fs}
\end{equation}
The coefficient $f_C$ has not been calculated using the MRE in the general case of massive quarks, but  it was shown in \cite{Madsen:1994vp} that the expression
\begin{equation}
f_{C,i}(k)=\frac{1}{12 \pi^{2}}\left[1-\frac{3 k}{2 m_i}\left(\frac{\pi}{2}-\arctan \frac{k}{m_i}\right)\right]
\label{eq:fc}
\end{equation}
has the right  limiting values for $m \rightarrow 0$ and $m \rightarrow \infty$, and is in excellent agreement with shell model calculations.

In the MRE framework, the thermodynamic integrals are obtained from the bulk ones by means of the following replacement \cite{Lugones:2018qgu,Lugones:2016ytl}:
\begin{equation}
\int_{0}^{\infty} \cdots \frac{k^{2} d k}{2 \pi^{2}} \longrightarrow \int_{\Lambda_i}^{\infty} \cdots \frac{k^{2} d k}{2 \pi^{2}} \rho_i(k) .
\label{eq:prescription}
\end{equation}
Since leptons form a uniform background, this prescription applies only to quarks, which feel the strong interaction and are confined within a finite region. The surface term for gluons is zero, as it is for massless particles.
To avoid unphysical negative values of $\rho_{i}(k)$ at small momenta, an infrared cutoff $\Lambda_i$ is introduced which is defined as the largest solution of the equation $\rho_i(k)=0$ with respect to  $k$.  $\Lambda_{i}$ depends on $m_i$ and $R$ and its values are given in Table I of Ref. \cite{Lugones:2020qll}.

The grand thermodynamic potential of the vector MIT bag model including finite size effects reads:
\begin{equation}
\begin{aligned}
\Omega V  =& - \sum_{i= q, l} \frac{g_{i} V}{6 \pi^{2}} \int_{\Lambda_{i}}^{\infty}  \frac{(f^*_{i+} + f^*_{i-} ) }{\sqrt{k^2 + m_{i} ^2}}  k^4 \rho_{i} \; d k    +  B V  \\
&- V \tfrac{1}{2} G_V \left(n^{\mathrm{MRE}}_u +  n^{\mathrm{MRE}}_d +  n^{\mathrm{MRE}}_s\right)^2,  
\end{aligned}    
\label{eq:Omega_MRE_full}
\end{equation}
being $n^{\mathrm{MRE}}_{q}$  the particle number density in the MRE formalism:
\begin{eqnarray}
n^{\mathrm{MRE}}_{q} =  \frac{g_{q}}{2 \pi^2} \int_{\Lambda_{q}}^{\infty} (f^*_{q+}- f^*_{q-}) k^2 \rho_{q}(k) dk.
\label{eq:number_MRE}
\end{eqnarray}

Replacing the MRE density of states given in Eq. \eqref{rho_MRE} into  Eq. \eqref{eq:number_MRE}, and separating volume, surface and curvature contributions we obtain:
\begin{equation}
n_{q}^{\mathrm{MRE}} =   n^V_q + \frac{S}{V} n^S_q + \frac{C}{V} n^C_q,    
\end{equation}
being $V=\tfrac{4}{3}\pi R^3$, $S=4\pi R ^2$, $C=8 \pi R$, and: 
\begin{eqnarray}
n^V_q  & \equiv &  \frac{g_{q}}{2 \pi^2} \int_{\Lambda_{q}}^{\infty}  (f^*_{q+}- f^*_{q-}) k^2 dk ,  \\
n^S_q  & \equiv &  g_{q}  \int_{\Lambda_{q}}^{\infty}  (f^*_{q+}- f^*_{q-}) f_{S, q} k d k ,  \label{eq:nS}  \\
n^C_q &  \equiv &  g_{q}  \int_{\Lambda_{q}}^{\infty}  (f^*_{q+}- f^*_{q-}) f_{C, q}  d k .   \label{eq:nC}
\end{eqnarray}
The quantity $n^V_q$ is always positive and represents the volume contribution to the particle number density. The surface contribution $n^S_q$ is always negative because $f_{S, q}<0$, which means that (for given $T$ and $\mu_q$) finite size effects reduce the particle number density with respect to the bulk case \footnote{Notice that, although the form of $f_{S,q}$ is model dependent, its sign can be understood intuitively because, in a finite system, the number of available states \textit{is reduced} with respect to the bulk due to quantization. This behavior is common  to MRE, finite box calculations and shell models, indicating that the effect encoded in $f_{S,i}$ is quite general.}. The curvature contribution $n^C_q$ has a more involved integrand but we have checked numerically that $n^C_q <0$ for all cases considered here.

Using the same procedure in Eq. \eqref{eq:Omega_MRE_full} we find:
\begin{equation}
\begin{aligned}
\Omega V = & - P V +  \sigma S + \gamma C \\
- & \sum_{i, j} 12 \pi G_V  \left (n^S_i n^C_j + n^S_j n^C_i +\frac{2}{R} n^C_i n^C_j  \right) .
\end{aligned}    
\label{eq:Omega_3}
\end{equation}
The coefficients of $V$,  $S$ and $C$ in Eq. \eqref{eq:Omega_3} are respectively the total pressure $P$, total surface tension $\sigma$ and total curvature energy $\gamma$ given by:
\begin{eqnarray}
P &= &  \sum_{q,l} P_i^{*}  + \tfrac{1}{2} \big(\sum_{q}  G_V^{1/2} n^V_q \big)^2 - B , \\
\sigma &= &  \sum_{q}  \sigma_{q}^{*} - \sum_{\substack{i=u,d,s \\ j=u,d,s}}  G_V n^V_i n^S_j , \label{eq:sigma_tot} \\
\gamma &=&   \sum_{q}   \gamma_{q}^{*} - \sum_{\substack{i=u,d,s \\ j=u,d,s}}  G_V n^V_i n^C_j  - \tfrac{3}{4} \big(\sum_{q}  G_V^{1/2} n^S_q \big)^2 , \quad  \label{eq:gamma_tot}
\end{eqnarray}
being
\begin{eqnarray}
P_i^{*} & = & \frac{g_i}{6 \pi^{2}} \int_{\Lambda_{i}}^{\infty} \frac{(f^*_{i+} + f^*_{i-} )}{\sqrt{k^2 + m_i ^2}}  k^4 d k ,  \label{eq:pressure_free} \\
\sigma_{i}^{*} & = &  - \frac{g_{i}}{3} \int_{\Lambda_{i}}^{\infty}   \frac{(f^*_{i+} + f^*_{i-}) f_{S, i} k^3 d k }{\sqrt{k^2 + m_i^2}}  ,  \label{eq:sigma_free} \\
\gamma_{i}^{*}  & = & 
- \frac{g_{i}}{3} \int_{\Lambda_{i}}^{\infty}   \frac{(f^*_{i+} + f^*_{i-} ) f_{C, i} k^2 d k }{\sqrt{k^2 + m_i^2}} .  \label{eq:gamma_free}  
\end{eqnarray}
The expressions for $P_i^{*}$,  $\sigma_i^{*}$ and $\gamma_i^{*}$ resemble respectively the pressure, the surface tension and the curvature energy of a non-interacting Fermi-Dirac gas, but they are calculated using the distribution functions $f^*_{i \pm}$ instead of $f_{i \pm}$ (see \cite{Lugones:2020qll} for comparison\footnote{Unfortunately, there was a missing overall minus sign in front of the integrals for $\sigma$ and $\gamma$ in Eqs. (10) and (11) of Ref. \cite{Lugones:2020qll}. That typo was not present in the calculations.}). Notice that $\sigma$ and $\gamma$ contain a ``free particle" term which is the sum of the contribution of each flavor, and new terms arising from vector interactions where all flavors are mixed. 
The ``free particle'' terms of $\sigma$ and $\gamma$ are positive because of the minus signs in Eqs. \eqref{eq:sigma_free} and \eqref{eq:gamma_free} and the behavior of $f_{S, q}$ and $f_{C, q}$.
The second term in Eq. \eqref{eq:sigma_tot} increases $\sigma$ because  $n^V_i >0$,  $n^S_j<0$, $G_V >0$, and there is an overall minus sign. The second term in Eq. \eqref{eq:gamma_tot} increases $\gamma$ because $n^C_j<0$, but the third term reduces it. 
Aside from these extra terms, vector interactions have an influence in $\sigma$ and $\gamma$  via the effective chemical potentials $\mu_{q}^{*}$ that enter the 
Fermi–Dirac distribution functions $f^*_{i \pm}$ (see Eqs. \eqref{eq:effective_mu} and \eqref{eq:effective_FD}). To determine $\mu_{q}^{*}$, the quantity $m_{V} V_{0}$ must be determined by solving self-consistently the equation for the vector field: 
\begin{equation}
m_{V} V_{0} = \sqrt{G_V} (n^{\mathrm{MRE}}_u + n^{\mathrm{MRE}}_d +  n^{\mathrm{MRE}}_s) .
\label{eq:gap_equation}
\end{equation}

In the present model, $\sigma$ and $\gamma$ are independent of the bag constant because $B$ is absorbed in the pressure $P$. 
Finally, note that the terms in the last line of Eq. \eqref{eq:Omega_3} are of order $R^0$ or $R^{-1}$, i.e. much smaller than the volume, surface and curvature terms, having a  negligible contribution to the thermodynamic potential.

\section{Results and Conclusions}
\label{sec:4}

\begin{figure*}[tb]
\centering
\includegraphics[angle=0,scale=0.38]{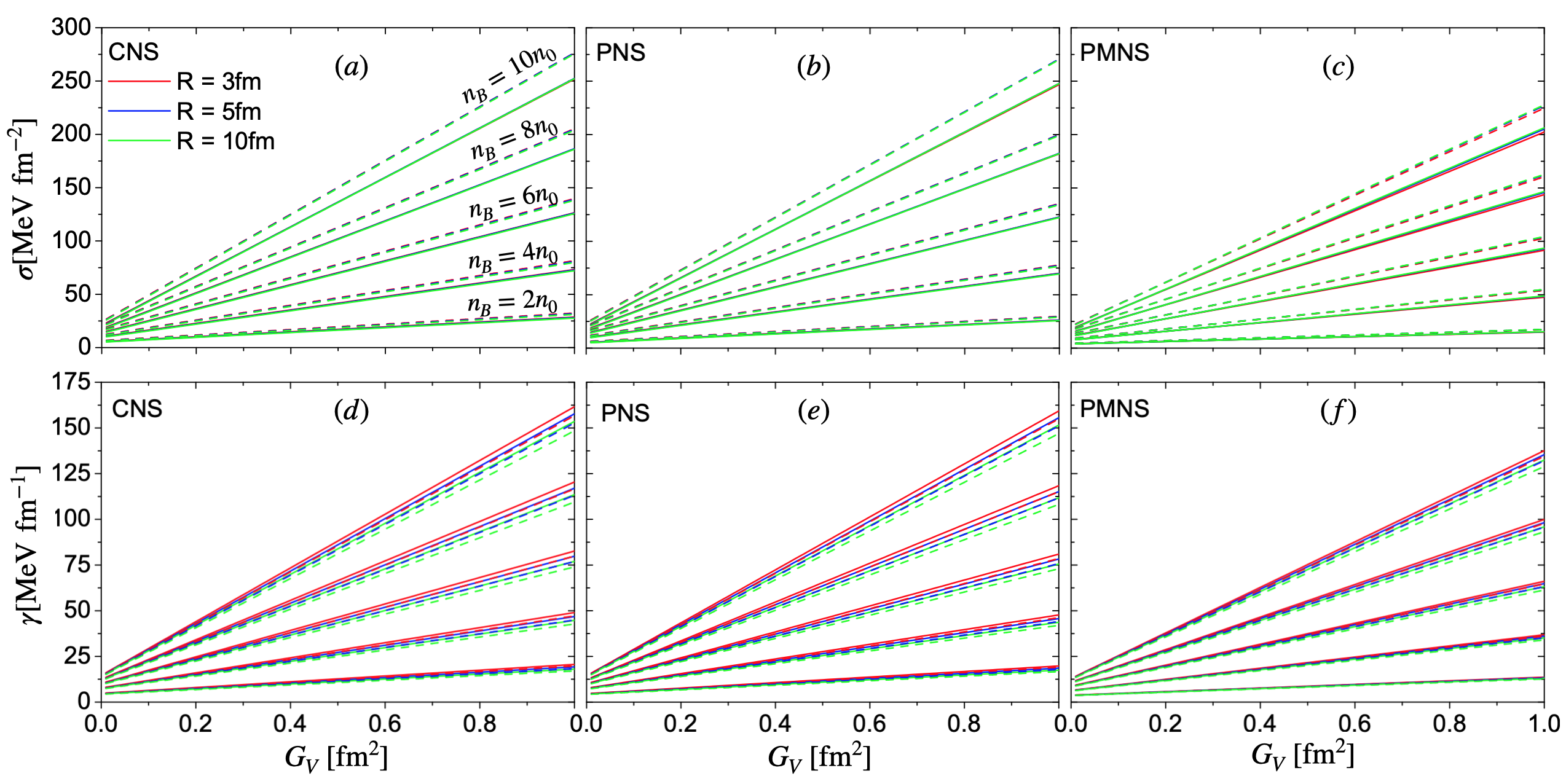}
\caption{Surface tension and curvature energy as a function of $G_V$ for different values of $n_B/n_0$. Each panel represents an astrophysical scenario introduced in Sec. \ref{sec:4}. Solid and dashed lines correspond to $\xi = 0$ and $\xi = -0.5$ respectively.}
\label{fig:1}
\end{figure*}

\begin{figure*}[tb]
\centering
\includegraphics[angle=0,scale=0.38]{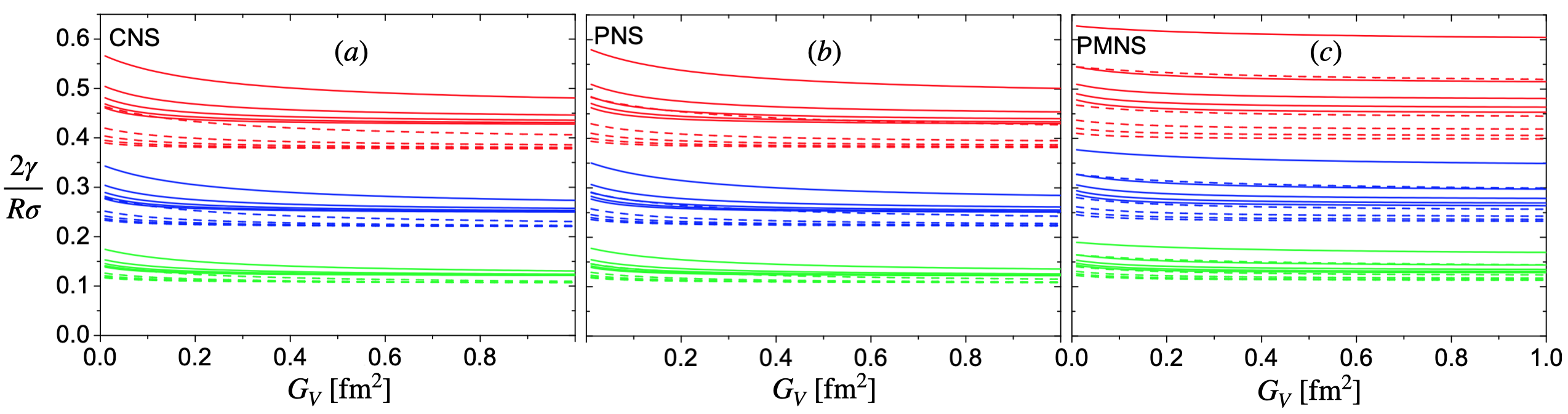}
\caption{Ratio of the curvature and the surface terms in the grand thermodynamic potential as a function of $G_V$ for different values of $n_B$. Colors and lines have the same meaning as in Fig. \ref{fig:1}. Within each type of curve,  $n_B$ increases from top to bottom.} 
\label{fig:2}
\end{figure*}

The mixed phase contains quark matter droplets in chemical equilibrium under weak interactions, which means that the chemical potentials of different species ($u$, $d$, and $s$ quarks, $e^{-}$ and $\nu_e$) are related by 
\begin{eqnarray}
\mu_d &=& \mu_u + \mu_e - \mu_{\nu_e},  \label{eq:chenical_1} \\
\mu_s &=& \mu_d  .    \label{eq:chenical_2}
\end{eqnarray}

In a self consistent analysis of the mixed phase, the electric charge density $n_Q$ and the Coulomb potential $\phi$ should be determined at each position of the quark droplet by solving the Poisson equation, as done for example in Refs. \cite{Voskresensky:2002hu,Alford:2006bx,Maslov:2018ghi,Yasutake:2014oxa,Wu:2017xaz,Wu:2018zoe}.  However, to keep our analysis as general as possible, we  will consider $n_Q$ and $\phi$ as free inputs in order to assess the dependence of $\sigma$ and $\gamma$ on that quantities. 
Since $\sigma$ and $\gamma$ are determined by the state of quark  matter at the droplet's inner boundary, we will focus on $n_Q$ and $\phi$ in that region.

The electron background in the mixed phase is uniform, i.e. it is the same at the internal and the external side of the drop's boundary. Additionally, the hadronic component is predominantly positive since it is constituted mainly of neutrons and protons.  Therefore, due to global charge neutrality, the  quark phase inside the drop has to be negative (c.f. Figs. 5 and 6 of Ref. \cite{Endo:2005zt}). Thus, $n_Q$ must be negative at the inner side of the drop:
\begin{eqnarray}
n_Q  \equiv \left( \tfrac{2}{3} n_u^{\mathrm{MRE}} - \tfrac{1}{3} n_d^{\mathrm{MRE}} - \tfrac{1}{3} n_s^{\mathrm{MRE}} - n_e\right) \le  0.
\label{eq:global_charge}
\end{eqnarray}
For convenience, we will write the charge density in terms of the charge-per-baryon ratio:  
\begin{equation}
\xi \equiv \frac{n_Q}{n_B},   
\end{equation}
where $n_B$ is the baryon number density and $\xi \leq 0$. 
From our calculations we learn that  $\xi$ must be  $\gtrsim -2$ for obtaining  $n_u^{\mathrm{MRE}} >0$. Thus, the expected values of $\xi$ are in the range $-2< \xi \leq 0$.

The Coulomb potential $\phi$ at the inner side of the drop's surface is determined by the charge enclosed within it. Since this charge is negative, the resulting electrostatic interaction is  repulsive on the electrons and $d$, $s$ quarks and attractive on $u$ quarks. Calculations show that $\phi$ is in fact negative and takes values between $0$ and $-50 \, \mathrm{MeV}$ at the drop's boundary (see e.g. Fig. 4 of Ref. \cite{Yasutake:2014oxa}). However, notice that Eqs. \eqref{eq:gap_equation}-\eqref{eq:global_charge}, are invariant by the change of variables $\mu_i \longrightarrow \mu_i + q_q |e \phi|$. As a consequence, $\sigma$ and $\gamma$ are independent of the value of $\phi$, as can be checked numerically.

We focus here in thermodynamic conditions that are representative of the following astrophysical scenarios:

(1) \textit{Cold deleptonized NSs (CNS)}, characterized by very low temperatures (below $\sim 1 \, \mathrm{MeV}$) and no trapped neutrinos. We adopt here $T=1 \, \mathrm{MeV}$ and $\mu_{\nu_e}=0$.  

(2) \textit{Hot lepton rich proto NSs (PNS)}, with temperatures up to $\sim 40$ MeV and a large amount of trapped neutrinos. As a representative case we consider here $T= 30$ MeV and $\mu_{\nu_e} = 100$ MeV \cite{Fischer:2017lag,Fischer:2021tvv}. 

(3) \textit{Postmerger NSs (PMNS)} According to numerical simulations,  the just merged compact object may attain temperatures up to $100 ~\mathrm{MeV}$ \cite{Most:2018eaw,Weih:2019xvw,Bauswein:2018bma,Bauswein:2020aag}, and huge neutrino trapping can be expected. As a limiting case we adopt $T= 100 ~\mathrm{MeV}$  and $\mu_{\nu_e} = 200 ~\mathrm{MeV}$.

In Fig. \ref{fig:1} we show $\sigma$ and $\gamma$ as  functions of $G_V$ for five values of $n_B$, namely $2 n_0$, $4 n_0$, $6  n_0$, $8  n_0$ and $10 n_0$, being $n_0=0.16 \mathrm{fm^{-3}}$ the nuclear saturation density.  As a general feature for all astrophysical scenarios, $\sigma$ and $\gamma$ grow linearly with $G_V$ and the slope increases with the baryon number density. The resulting effect is an increase of $\sigma$ and $\gamma$ by a factor of $\sim 10$ when $G_V$ grows from $0$ to $1 \, \mathrm{fm}^2$ (same range of values as in \cite{Salinas:2019fmu,Lopes2021_II}). Negatively charged droplets have larger $\sigma$ and smaller $\gamma$ than charge neutral ones, their difference being around $10 \%$ for $\xi =-0.5$ as taken here. More negative values of $\xi$ do not affect significantly $\sigma$ and $\gamma$. 
Notice that for the particular case of $G_V=0$, $\sigma$ and $\gamma$ grow considerably with $n_B$  as shown in previous works  for a variety of astrophysical conditions \cite{Lugones:2016ytl,Lugones:2018qgu,Lugones:2020qll}. The density dependence is more pronounced when vector interactions are turned on, since they also raise with $n_B$.
Finally, the combined effect of large temperatures and neutrino trapping produces a decrease of $\sigma$ and $\gamma$ as seen in Figs. \ref{fig:1}c and \ref{fig:1}f.

In Fig. \ref{fig:2} we show the ratio $\gamma C/(\sigma S ) = 2 \gamma /(R \sigma )$ between the surface and curvature terms in the grand thermodynamic potential (see Eq. \eqref{eq:Omega_3}) to evaluate the relative weight of each contribution. Curves are quite horizontal indicating that the ratio is insensitive to $G_V$. The relative importance of curvature with respect to  surface effects depends mostly on the droplet's radius, being around $10\%$ for $10 \mathrm{fm}$ and around $50\%$ for $3 \mathrm{fm}$. The ratio is larger for smaller densities, and smaller when droplets are negatively charged. The range of values of the ratio is wider for smaller radii and when temperatures and with neutrino trapping are increased. 

The main new conclusion of the above results is that 
$\sigma$ and $\gamma$ are strongly enhanced by vector interactions. Within the \textit{non-interacting} MIT bag model, we find typically  $\sigma \approx 2-25 ~\mathrm{MeV fm^{-2}}$ and  $\gamma \approx 2-12~ \mathrm{MeV fm^{-1}}$. When repulsive vector interactions are turned on these values are increased by a large factor, that can be as large as 10 for $n_B = 10 n_0$ and $G_V= 1 \, \mathrm{fm}^2$.
The effect of vector interactions on $\sigma$ and $\gamma$ can be understood qualitatively as follows. Surface tension in the MRE formalism is a quantum effect arising from the smaller number of states available in a finite system with respect to the bulk case. As a consequence, for a given number of particles, higher energy levels are occupied in a finite region than in the bulk, and the extra energy is interpreted as being stored in the surface and the curvature. Vector interactions shift the energy levels to even higher values, as seen from the dispersion relation $E_q=(m_q^2+k^2)^{1/2} +g V^0$  (notice that $g V^0 >0$), increasing not only the pressure but also  $\sigma$ and $\gamma$. As a result, the here found effect of vector interactions on $\sigma$ and $\gamma$ is fairly model independent.
This behavior has strong consequences on the internal structure of hybrid stars. With such large $\sigma$ and $\gamma$, the energy cost of surface and curvature makes more difficult to compensate the energy gain of global charge neutrality. As a consequence, we expect as a general feature that vector repulsive interactions will decrease the range of densities where the mixed phase is energetically favored, specially at large densities. In the extreme cases considered here a sharp interface could occur.

\bigskip

\subsection*{Acknowledgements}
G.L. acknowledges the support of the Brazilian agencies Conselho Nacional de Desenvolvimento Cient\'{\i}fico e Tecnol\'ogico (grant 309767/2017-2) and Funda{\c c}\~ao de Amparo \`a
Pesquisa do Estado de S\~ao Paulo  (grant 2013/10559-5).
A. G. G. would like to acknowledge to CONICET for financial support under Grant No. PIP17-700.  

\bibliography{vector_MIT}

\end{document}